\documentclass[showpacs,preprintnumbers]{revtex4}

\begin{document}

\title{Transport Coefficients of Non-Newtonian Fluid and Causal Dissipative
Hydrodynamics }
\author{T. Koide and T. Kodama}
\affiliation{Instituto de F\'{\i}sica, Universidade Federal do Rio de Janeiro, C. P.
68528, 21945-970, Rio de Janeiro, Brazil}

\begin{abstract}
A new formula to calculate the transport coefficients of the causal
dissipative hydrodynamics is derived by using the projection operator method
(Mori-Zwanzig formalism) in \cite{koide1}. This is an extension of the
Green-Kubo-Nakano (GKN) formula to the case of non-Newtonian fluids, which
is the essential factor to preserve the relativistic causality in
relativistic dissipative hydrodynamics. This formula is the generalization
of the GKN formula in the sense that it can reproduce the GKN formula in a
certain limit. In this work, we extend the previous work so as to apply to
more general situations.
\end{abstract}

\maketitle

\section{Introduction}

Hydrodynamic models have been extensively applied to analyze the collective
aspects of relativistic heavy-ion collisions. These analyses have mainly
been done so far for ideal fluids \cite{Hidros}. The effect of dissipation
(viscosity and heat conduction) to this problem has started only recently
and it is less well understood yet. One of the reasons for this is that,
besides technical difficulties in numerical implementations, a relativistic
extension of the dissipative hydrodynamics is not trivial at all
conceptually \cite{muller,ii,carter,OG,dkkm1,dkkm2,dkkm3,dkkm4,jou1}. A
naive covariant extension of the Navier-Stokes equation leads to the problem
of relativistic acausality and instabilities of the theory \cite%
{Instabilities,kouno,Gio,dkkm3}.

An essential factor to solve this problem is to introduce a memory effect
with a finite relaxation time in the definition of irreversible currents 
\cite{dkkm1,dkkm4,kkr,heat}. An important point here is that, with the
presence of memory effects, the fluid becomes \textit{non-Newtonian}, that
is, the irreversible current is not simply proportional to the
thermodynamical forces.

This raises several serious questions in applying the causal dissipative
hydrodynamics to various phenomena at relativistic energies. The crucial
one, we will focus in this paper, is that we cannot use the Geen-Kubo-Nakano
(GKN) formula to calculate the transport coefficients, because the
derivation depends on the Newtonian property of the fluid. See the
discussion in Appendix \ref{app:gkn}. To obtain the transport coefficients
of the causal dissipative hydrodynamics, a new formulation should be
developed.

One possible approach to obtain the transport coefficients in the presence
of memory effects is the so-called projection operator method (POM). The POM
was originally proposed to obtain master equations and generalized Langevin
equations from microscopic dynamics by implementing systematic
coarse-grainings in terms of projection operators for macroscopic variables 
\cite{nakajima,zwanzig,mori,review,fick,reichl,b-zwanzig,kubo}. It is also
known that the POM is useful to obtain the microscopic expressions of
various transport coefficients \cite{koide1,fick,reichl,b-zwanzig}. In the
POM, the transport coefficients are related to the memory function of the
generalized Langevin equation. Except for trivial cases, it is very
difficult to evaluate the memory function exactly and some appropriate
approximations are needed. A common method is to neglect a part of the
projection operator in the memory function (as explained later). In the
following, we call such an approximation as the \textit{Q approximation}. It
is known that the formula for the transport coefficients obtained with the Q
approximation in the POM are equivalent to those of the GKN formula \cite%
{koide1,fick,reichl,b-zwanzig}. It is further known that the coarse-grained
equation of a conserved quantity obtained by the POM with the Q
approximation becomes a usual diffusion equation \cite%
{fick,reichl,b-zwanzig,koide6}. That is, the use of the Q approximation in
the POM leads to the behaviors of Newtonian fluids.

Recently, one of the present authors discussed the coarse-graining procedure
in the POM without using the Q approximation \cite{km,koide6,koide1}. There,
it was shown that the equation for a conserved number becomes a
telegraph-type equation when the Q approximation is not introduced\cite%
{koide6}. Note that the telegraphic equation is derived when a memory effect
is introduced in a diffusion equation \cite{koide6}. This indicates that we
can apply this method to construct the causal dissipative hydrodynamics in
the POM, defining the microscopic expressions of the transport coefficients
for non-Newtonian fluids in a consistent manner. Following this idea, new
formulas of the transport coefficients for the causal dissipative
hydrodynamics have been derived \cite{koide1}. This new formula differs in a
essential way from those obtained using the GKN formula with the Newtonian
case, although it can reproduce the GKN formula under a limit where the Q
approximation is valid.

In this paper, we present a more detailed version of the work of \cite%
{koide1} and derive more general expressions of transport coefficients, in
particular, the shear viscosity of the causal dissipative hydrodynamics.
This paper is organized as follows. In section \ref{chap:pom}, for the sake
of later convenience, we review briefly the projection operator method to
derive the generalized Langevin equation. In section \ref{chap:memory}, the
so-called Mori projection operator is introduced. We calculate explicitly
the memory function in section \ref{chap:reex}. This result is the
generalization of the formula obtained in \cite{koide1} and one of the main
results of this paper. By using this general expression, we define the
causal shear viscosity coefficient and the relaxation time in section \ref%
{chap:n1shear}. The result of this section is completely same as that of 
\cite{koide1}. The relation between our formula and the GKN formula is
discussed in section \ref{chap:gkn}. In section \ref{chap:n2mct}, we apply
the the result to an exactly solvable model to confirm the validity of our
exact expression of the memory function. In section \ref{chap:n2shear}, we
reinvestigate the result of the section \ref{chap:n1shear} and propose
another possible definition of the causal shear viscosity coefficient. The
section \ref{chap:dis} is devoted to concluding remarks.

\section{Projection Operator Method}

\label{chap:pom}

It should be emphasized that the projection operator method (POM) was
firstly proposed by Nakajima \cite{nakajima}, although it is often refereed
as the Mori-Zwanzig formalism due to the extensive use and developments done
by these authors. This approach has been studied so far in various contexts
of physics and chemistry \cite{fick,reichl,b-zwanzig}. In particular, Mori
introduced the so-called Mori projection operator to describe the dynamics
near thermal equilibrium and derived a generalized Langevin equation from
microscopic models \cite{mori}. The generalized Langevin equation (the Mori
equation) gives the basis of the various development of statistical physics.
Kawasaki, for example, developed the mode coupling theory which describes
the dynamical critical phenomena by using the technique of the generalized
Langevin equation \cite{kawasaki1}. The mode coupling theory is recently
used to discuss the glass dynamics \cite{mct}. It is considered that the POM
is a promising method to establish a new coarse-grained dynamics like the
dynamical density functional theory \cite{yoshimori,kawasaki2}. The
formulation of the projection operator method has been polished up by
several authors \cite%
{shibata1,shibata2,shibata3,shibata4,koide2,koide3,koide4,koide5}. The
derivation discussed here is following \cite{koide3}.

In a quantum mechanical system, the time evolution of an operator is
governed by the Heisenberg equation of motion, 
\begin{eqnarray}
\frac{d}{dt}O(t) &=&i[H,O(t)]  \nonumber \\
&=&iLO(t) \\
\longrightarrow O(t) &=&e^{iL(t-t_{0})}O(t_{0}),  \label{eqn:HE}
\end{eqnarray}%
where $L$ is the Liouville operator and $t_{0}$ is an initial time at which
we prepare an initial state. In the following, we set $t_{0}=0$. We consider
here an isolated system so that the Hamiltonian is independent of time. Note
that Eq. (\ref{eqn:HE}) is also valid for classical cases provided that the
commutator of the Liouville operator is interpreted as the Poisson bracket.

In order to derive coarse-grained equations such as hydrodynamical equation
of motion from a microscopic theory, we should construct a closed system of
equations expressed only by those variables with macroscopic properties of
the system. However, the Heisenberg equation of motion contains the
information not only of gross variables associated with macroscopic
(hydrodynamic) time scales, but also of microscopic variables. In the POM,
the latter variables are projected out by introducing an appropriate
projection operator $P$ (to be specified later). We denote its complementary
operator by $Q(=1-P).$ They should satisfy, 
\begin{eqnarray}
P^{2} &=&P, \\
PQ &=&QP=0.
\end{eqnarray}%
Here, the projection operators are time-independent. To describe real
non-equilibrium processes, in general, the projection operator should be
time-dependent. However, for the purpose of the present paper, simple
time-independent projection operators are suffice as the definition of the
transport coefficients of the relativistic dissipative fluid.

From Eq. (\ref{eqn:HE}), one can see that the time dependence of operators
is determined by $e^{iLt}$. This operator obeys the following differential
equations, 
\begin{equation}
\frac{d}{dt}e^{iLt}=e^{iLt}iL=e^{iLt}(P+Q)iL.  \label{eqn:P+Q}
\end{equation}%
Multiplying the operator $Q$ from the right, we have 
\begin{equation}
\frac{d}{dt}e^{iLt}Q=e^{iLt}PiLQ+e^{iLt}QiLQ.  \label{eqn:Q}
\end{equation}%
Equation (\ref{eqn:Q}) can be solved for $e^{iLt}Q$, 
\begin{equation}
e^{iLt}Q=Qe^{iLQt}+\int_{0}^{t}d\tau e^{iL\tau }PiLQe^{iLQ(t-\tau )}.
\label{eqn:TC-dainyuu}
\end{equation}%
Substituting Eq. (\ref{eqn:TC-dainyuu}) into the last term in Eq. (\ref%
{eqn:P+Q}) and operating $O(0)$ from the right, we obtain the so-called
time-convolution (TC) equation, 
\begin{eqnarray}
\frac{d}{dt}O(t) &=&e^{iLt}PiLO(0)+\int_{0}^{t}d\tau e^{iL(t-\tau
)}PiLQe^{iLQ\tau }iLO(0)  \nonumber \\
&&+Qe^{iLQt}iLO(0).  \label{eqn:TC-1}
\end{eqnarray}%
The first term on the r.h.s. of the equation is called the streaming term
and usually corresponds to collective modes such as plasma wave, spin wave
and so on. The second term is the memory term that causes dissipation. The
third term represents the noise term. The second term and third terms are
related through the fluctuation-dissipation theorem of second kind, which
will be discussed later.

Discussion of this section has been done in the Heisenberg picture and the
generalized Langevin equation (\ref{eqn:TC-1}) is derived. We can develop
the similar discussion in the Schr\"{o}dinger picture and obtain master
equations. As we pointed out, to discuss more complex non-equilibrium
processes, we have to change the basis of the projection with time. Then,
the projection operator is explicitly time-dependent. As for the operation
of the time-dependent projection operator, see \cite{koide4} and references
therein.

It is also possible to derive another form of the generalized Langevin
equation, which is called the time-convolutionless (TCL) equation. There are
some cases where we cannot implement the Markov approximation in the TC
equation. The $\phi^4$ theory is one of the examples, and the Markov
equation is derived from the TCL equation. See \cite{koide5} for details.

\section{Mori Projection Operator}

\label{chap:memory}

In the above derivation of the TC equation, we have not specified the
projection operator $P$. As a matter of fact, there are many possible
projection operators that extract the slowly varying components from
dynamics. Suppose that the macroscopic dynamics can be described by the time
evolutions of $n$-gross variables. Then we have to define the projection
operator to project any time evolution onto the space spanned by these $n$%
-gross variables. For example, in the case of usual hydrodynamics, the time
evolutions are described by the dynamics of the energy density, velocity
field and number density. Then we have five variables which form a complete
set for hydrodynamics (two scalar fields and one vector field).

Strictly speaking, there is no general criterion to prepare a complete set
of gross variables. It is, however, suggested that there are three
candidates for gross variables \cite{mori,kawasaki1} : (i) order parameters
(if there exists phase transitions) , (ii) density variables of conserved
quantities and (iii) their products. Once we find out a complete set of the
gross variables, any macroscopic variables should be approximately given by
a linear combination of these gross variables (see below). We can use the
Mori projection operator to implement this coarse-graining.

Let us represents a set of gross variables by a $n$-dimensional vector, 
\begin{equation}
\mathbf{A}^{T}=(A_{1},A_{2},\cdots ,A_{n}).  \label{pro-mori}
\end{equation}%
Then, the Mori projection operator $P$ is defined as 
\begin{eqnarray}
P~O=\sum_{i=1,}^{n}c_{i}A_{i},
\end{eqnarray}
for an arbitrary operator $O,$ where the coefficient $c_{i}$ is given by 
\begin{eqnarray}
c_{i}=\sum_{j=1}^{n}(O,A_{j}^{\dagger })\cdot (A,A^{\dagger })_{ji}^{-1}.
\end{eqnarray}
The inner product is Kubo's canonical correlation, 
\begin{eqnarray}
(X,Y)=\int_{0}^{\beta }\frac{d\lambda }{\beta }\mathrm{Tr}[\rho ~e^{\lambda
H}Xe^{-\lambda H}Y],  \label{eqn:kubo}
\end{eqnarray}
where $\rho =e^{-\beta H}/\mathrm{Tr}[e^{-\beta H}]$ with the temperature $%
\beta ^{-1}$. The inverse of the canonical correlation is defined by 
\begin{eqnarray}
\sum_{j}(A,A)_{ij}^{-1}\cdot (A_{j},A_{k})=\delta _{i,k}.
\end{eqnarray}

As for the physical meaning of the Mori projection operator, see, for
example, \cite{mori,km}.

\section{The exact expression of the Memory Function}

\label{chap:reex}

Substituting the Mori projection operator into Eq. (\ref{eqn:TC-1}), the TC
equation is reexpressed as follows; 
\begin{eqnarray}
\frac{\partial }{\partial t}\mathbf{A}(t)=i\Delta \mathbf{A}%
(t)-\int_{0}^{t}d\tau \mathbf{\Xi }(\tau )\mathbf{A}(t-\tau )+\xi (t),
\end{eqnarray}
where 
\begin{eqnarray}
i\Delta &=&\sum_{k}(iLA_{i},A_{k}^{\dagger })(A,A^{\dagger })_{kj}^{-1}, \\
\mathbf{\Xi }_{ij}(t) &=&-\theta
(t)\sum_{k}(iLQe^{iLQt}iLA_{i},A_{k}^{\dagger })(A,A^{\dagger })_{kj}^{-1},
\label{xi} \\
{\xi }_{i}(t) &=&Qe^{iLQt}iLA_{i}.
\end{eqnarray}%
The TC equation using the Mori projection operator is called the Mori
equation.

The important information of dissipation is given by the memory function $%
\mathbf{\Xi }(t)$. As a matter of fact, transport coefficients are defined
by it. However, the calculation of the memory function is not simple because
the expression of the memory function has the projection operator $Q$ (see
Eq. (\ref{xi})). As a matter of fact, the projection operator $Q$ is
approximately replaced with $1$ to estimate the memory function in many
textbooks \cite{b-zwanzig,fick,reichl}, 
\begin{eqnarray}
\mathbf{\Xi }(t)\approx -\theta (t)\sum_{k}(iLe^{iL\tau
}iLA_{i},A_{k}^{\dagger })(A,A^{\dagger })_{kj}^{-1}.
\end{eqnarray}
As mentioned in the introduction, we call this procedure the Q approximation.

Recently, Okada et al. calculated the memory function of the Ising model and
found that the memory function can be expressed in terms of the combination
of the usual time correlation functions \cite{okada}. Afterwords, Koide
applied the same procedure to microscopic models and discussed the effect of
the Q approximation \cite{km,koide6}. The coarse-grained dynamics of
conserved quantities of the model are, usually, considered to be given by
the diffusion equation. As a matter of fact, when we apply the Q
approximation, we can derive the diffusion equation in the model. When we do
not apply the Q approximation, however, the coarse-grained dynamics is given
by the telegraph-type equation instead of the diffusion equation \cite%
{koide6}. More interestingly, if the model has a conserved quantity, we can
derive the sum rule associated with the conserved law. It was shown that the
telegraph-type equation derived with the memory effect is consistent with
the sum rule, while the diffusion equation (Q approximation) breaks it \cite%
{kadanoff,koide6}.

The same idea is used to define the transport coefficients of the causal
dissipative hydrodynamics by using the \textit{simplest} Mori projection
operator that is defined with only one gross variable \cite{koide1}. In this
section, we extend the discussion of \cite{koide1} to more complex cases
where the Mori projection operator is defined with $n$-gross variables.

To calculate the coarse-grained time-evolution operator, we introduce the
following operator, 
\begin{eqnarray}
\mathcal{B}(t) = 1 + \sum^{\infty}_{n=1}(-i)^n \int^{t}_{0}dt_1 \cdots
\int^{t_{n-1}}_{0}dt_{n} \breve{L}^P(t_1)\breve{L}^P(t_{2}) \cdots \breve{L}%
^P(t_{n}),  \label{B} \\
\end{eqnarray}
with 
\begin{eqnarray}
\breve{L}^P (t) &\equiv& e^{-iL t}PL e^{iL t}.
\end{eqnarray}
Then, the matrix $\mathbf{\Xi}(t)$ is rewritten as 
\begin{eqnarray}
\mathbf{\Xi}_{ij}(t) &=& -\theta(t) (iLe^{iLt}\mathcal{B}(t)QiLA_i,A^{%
\dagger}_j)(A_j,A_j^{\dagger})^{-1}  \nonumber \\
&=& -\theta(t) \left[ (\ddot{\mathbf{X}}(t))_{ij} - (\dot{\mathbf{X}}(0)\dot{%
\mathbf{X}}(t))_{ij} \right.  \nonumber \\
&& + \sum_{n=1}^{\infty}(-1)^n \int^{t}_{0}dt_1 \cdots
\int^{t_{n-1}}_{0}dt_n ( \ddot{\mathbf{X}}(t_n)\dot{\mathbf{X}}%
(t_{n-1}-t_n)\cdots \dot{\mathbf{X}}(t_1 - t_2)\dot{\mathbf{X}}(t-t_1) )_{ij}
\nonumber \\
&& \left. - \sum_{n=1}^{\infty}(-1)^n \int^{t}_{0}dt_1 \cdots
\int^{t_{n-1}}_{0} dt_n ( \dot{\mathbf{X}}(0)\dot{\mathbf{X}}(t_{n})\dot{%
\mathbf{X}}(t_{n-1}-t_n)\cdots \dot{\mathbf{X}}(t_1-t_2)\dot{\mathbf{X}}%
(t-t_1) )_{ij} \right].  \label{matrix-xi}
\end{eqnarray}
See Appendix \ref{app:2} for detailed discussion. By using the Laplace
transform, the above expression can be rewritten as the following simple
form, 
\begin{eqnarray}
\mathbf{\Xi}^L(s) = - \ddot{\mathbf{X}}^L(s)\frac{1}{1+\dot{\mathbf{X}}^L(s)}
+ \dot{\mathbf{X}}(0)\dot{\mathbf{X}}^L(s)\frac{1}{1+\dot{\mathbf{X}}^L(s)},
\label{eqn:mf}
\end{eqnarray}
where the functions $\dot{\mathbf{X}}^L(s)$ and $\ddot{\mathbf{X}}^L(s)$ are
given by the Laplace transform of the following correlation functions, 
\begin{eqnarray}
\dot{\mathbf{X}}_{ij}(t) &=& \sum_{k}(iLA_i
(t),A^{\dagger}_k)(A,A^{\dagger})^{-1}_{kj},  \nonumber \\
\ddot{\mathbf{X}}_{ij}(t) &=& \sum_{k}((iL)^2 A_i
(t),A^{\dagger}_k)(A,A^{\dagger})^{-1}_{kj}.
\end{eqnarray}
This is the exact expression of the memory function without using the Q
approximation. One can see that if we set $n=1$, the expression (\ref{eqn:mf}%
) reproduces the result of the previous result of \cite{koide1}. The
expression of the transport coefficients are derived by employing
approximations to this memory function, as will will see in the next section.


\section{Shear Viscosity of Causal dissipative Hydrodynamics in n=1 Form}

\label{chap:n1shear}

We apply the formula to define the shear viscosity coefficient of causal
dissipative hydrodynamics. For simplicity, we consider a particular case of
shear flow, where the fluid velocity points in the $x$ direction and varies
spatially in the $y$ direction \cite{b-zwanzig,koide1}. Then, the
energy-momentum tensor obeys the following equation of motion, 
\begin{eqnarray}
\frac{\partial}{\partial t}T^{0x}(y,t) = - \frac{\partial}{\partial y}%
T^{yx}(y,t) = - \frac{\partial}{\partial y}\pi^{yx}(y,t),
\end{eqnarray}
where $\pi^{\alpha\beta}$ is the traceless part of the energy-momentum
tensor, 
\begin{eqnarray}
\pi^{kl} = \left( \delta^k_i \delta^l_j - \frac{1}{3}\delta^{kl}\delta_{ij}
\right)T^{ij}~~~~~(i,j,k,l=1,2,3)
\end{eqnarray}

To define the shear viscosity coefficient, we introduce the Fourier
transform of $T^{0x}(y,t)$, and set $O(0)=T^{0x}(k_{y},0)$. And following
the usual derivation of the shear viscosity coefficient of the GKN formula,
we choose $T^{0x}(k_{y},0)$ as a unique gross variable \cite%
{koide1,b-zwanzig}. In this case, the Mori projection operator is defined by 
\begin{eqnarray}
PO=(O,T^{0x}(-k_{y},0))(T^{0x}(k_{y},0),T^{0x}(-k_{y},0))^{-1},
\end{eqnarray}
where $O$ is an arbitrary operator. We will discuss later a more involved
case where we need two gross variables to define the Mori projection
operator.

Then, the TC equation (\ref{eqn:TC-1}) is given by 
\begin{eqnarray}
\frac{\partial}{\partial t}T^{0x}(k_y,t) = i\Delta(k_y) T^{0x}(k_y,t) -
\int^t_0d\tau \Xi(k_y,t-\tau)T^{0x}(k_y,\tau) + \xi(k_y,t),
\label{eqn:pro-hydro}
\end{eqnarray}
where 
\begin{eqnarray}
i\Delta(k_y) &=&
(iLT^{0x}(k_y,0),T^{0x}(-k_y,0))(T^{0x}(k_y,0),T^{0x}(-k_y,0))^{-1}, \\
\Xi(k_y,t) &=& \frac{1}{2\pi i}\int_{\mathbf{Br}}\Xi^L(k_y,s)e^{st}ds, \\
\xi(k_y,t) &=& Qe^{iLQt}iLT^{0x}(k_y,t).
\end{eqnarray}
The Laplace transform of the memory function $\Xi^L(k_y,s)$ is given by 
\begin{eqnarray}
\Xi^L(k_y,s) &=& -\ddot{X}^L(k_y,s)\frac{1}{1+\dot{X}^L(k_y,s)},
\label{eqn:shearmemory}
\end{eqnarray}
where 
\begin{eqnarray}
\ddot{X}(k_y,t) &=& (iLT^{0x}
(k_y,t),T^{0x}(-k_y,0))(T^{0x}(k_y,0),T^{0x}(-k_y,0))^{-1},  \nonumber \\
\ddot{X}_{ij}(k_y,t) &=& ((iL)^2 T^{0x}
(k_y,t),T^{0x}(-k_y,0))(T^{0x}(k_y,0),T^{0x}(-k_y,0))^{-1}.
\end{eqnarray}
In this derivation, we used 
\begin{eqnarray}
\dot{X}(0) = (iLT^{0x}(k_y,0),T^{0x}(-k_y,0)) = 0.
\end{eqnarray}

So far, everything is exact formally. Now we carry out the coarse-grainings
of the time scale to break the time-reversal symmetry. For this purpose,
first of all, we separate the memory function into the two terms as follows 
\cite{koide1,koide6,km}, 
\begin{eqnarray}
\frac{\partial }{\partial t}T^{0x}(k_{y},t)=-\int_{0}^{t}d\tau \Omega
^{2}(k_{y},t-\tau )T^{0x}(k_{y},\tau )-\int_{0}^{t}d\tau \Phi (k_{y},t-\tau
)T^{0x}(k_{y},\tau ).
\end{eqnarray}
Here, we dropped the noise term. The frequency function and the renormalized
memory function are defined by 
\begin{eqnarray}
\Omega ^{2}(k_{y},t) &=&i\int \frac{d\omega }{2\pi }\mathrm{Im}[\Xi
^{L}(k_{y},-i\omega +\epsilon )]e^{-i\omega t}, \\
\Phi (k_{y},t) &=&\int \frac{d\omega }{2\pi }\mathrm{Re}[\Xi
^{L}(k_{y},-i\omega +\epsilon )]e^{-i\omega t},
\end{eqnarray}%
respectively.

To introduce the coarse-graining in time, we have to know the temporal
behavior of the two functions. The behavior of the two functions have been
investigated for some special cases, such as the chiral order parameter in
the Nambu-Jona-Lasinio model \cite{km}, exactly solvable model of many
harmonic oscillators \cite{km} and the non-relativistic model with a
conserved density \cite{koide6}. For all these cases, the two functions
exhibit common properties; the frequency function converges to a finite
value and the renormalized memory function vanishes at late time. Inspired
by these examples, we introduce an important assumption that these features
for the temporal behavior of the two functions are valid in general. That
is, the renormalized memory function relaxes rapidly and vanishes at large $t
$, while the frequency function converges to a finite value after short time
evolution. In principle, the validity of the assumption should be checked
for more general examples by implementing numerical calculations. Once we
accept the above assumption, we may introduce the following ansatzs for the
memory functions incorporating these basic features essentially \cite{km}: 
\begin{eqnarray}
\Omega ^{2}(k_{y},t)\longrightarrow D_{k_{y}}{k_{y}}^{2},~~~~\Phi
(k_{y},t)\longrightarrow \frac{2}{\tau _{k_{y}}}\delta (t),
\label{eqn:condition}
\end{eqnarray}
where 
\begin{eqnarray}
D_{k_{y}} &=&\frac{1}{{k_{y}}^{2}}\lim_{t\rightarrow \infty }\Omega
^{2}(k_{y},t),  \label{cdc} \\
\frac{1}{\tau _{k_{y}}} &=&\int_{0}^{\infty }dt\Phi (k_{y},t).  \label{rt}
\end{eqnarray}
The factor $k_{y}^{2}$ is introduced for the later convenience (see Eq.(\ref%
{T0x(yk)})). The above ansatzs are shown to be consistent with the final
value theorem of the Laplace transformation \cite{koide1,koide6,km}. That
is, when the renormalized memory function converges to zero at late time,
its Laplace transform $\Phi ^{L}(k_{y},s)$ should satisfy \cite{footnote2}, 
\begin{eqnarray}
\lim_{t\rightarrow \infty }\Phi (k_{y},t)=\lim_{s\rightarrow 0}s\Phi
^{L}(k_{y},s)=0.
\end{eqnarray}
Similarly, for the frequency function, 
\begin{eqnarray}
D_{k_{y}} k_{y}^{2}=\lim_{t\rightarrow \infty }\Omega
^{2}(k_{y},t)=\lim_{s\rightarrow 0}s(\Omega ^{2})^{L}(k_{y},s).
\end{eqnarray}

Using these expressions, we have the equation for the energy momentum tensor
component, 
\begin{equation}
\frac{\partial }{\partial t}T^{0x}(k_{y},t)=-D_{k_{y}}
k_{y}^{2}\int_{0}^{t}d\tau wu^{x}(k_{y})-\frac{1}{\tau _{k_{y}}}%
T^{0x}(k_{y},t).  \label{T0x(yk)}
\end{equation}
Here, we expressed the $x$-component of the fluid velocity as%
\begin{eqnarray}
u^{x}(k_{y})=T^{0x}(k_{y})/w,
\end{eqnarray}
where $w$ is an enthalpy density \cite{koide1}.

On the other hand, as pointed out before, the time evolution of the
energy-momentum tensor in causal dissipative hydrodynamics is given by a
kind of the telegraph equation . In particular, in the special case
discussed here ( the fluid velocity points in the $x$ direction and varies
spatially in the $y$ direction), the linearized equation of the causal
dissipative hydrodynamics is given by the following telegraph equation \cite%
{dkkm1,dkkm4}, 
\begin{eqnarray}
\frac{\partial ^{2}}{\partial t^{2}}T^{0x}+\frac{1}{\tau }\frac{\partial }{%
\partial t}T^{0x}+\frac{\eta ^{NN}}{2\tau }\frac{\partial ^{2}}{\partial
y^{2}}u^{x}=0.  \label{eqn:c-hydro}
\end{eqnarray}

This equation defines the causal shear viscosity coefficient $\eta^{NN}$ and
corresponding relaxation time $\tau$. By comparing Eq. (\ref{T0x(yk)}) with
Eq. (\ref{eqn:c-hydro}), we obtain the expression for the causal shear
viscosity coefficient and the respective relaxation time as 
\begin{eqnarray}
\eta ^{NN} &=&\lim_{k_{y}\rightarrow 0}2wD_{k_{y}}\tau .  \label{eqn:shear1}
\\
\tau &=&\lim_{k_{y}\rightarrow 0}\tau _{k_{y}},  \label{eqn:rela1}
\end{eqnarray}%
which are the results obtained in \cite{koide1}. In this derivation, it is
assumed that the projection operator is defined by only one gross variable.
We will reconsider this derivation in section \ref{chap:n2shear}.

\section{Green-Kubo-Nakano formula}

\label{chap:gkn}

We obtained an expression of the causal shear viscosity coefficient for
non-Newtonian fluids. On the other hand, it is well-known that the shear
viscosity coefficient of Newtonian fluids is given by the GKN formula. In
this section, we discuss the relation between our formula and the GKN
formula \cite{koide1}.

As was mentioned before, it is known that, when we apply the Q
approximation, the GKN formula is reproduced in the projection operator
method \cite{footnote1}. When we apply the Q approximation, the memory
function is given by \cite{b-zwanzig,fick,reichl}, 
\begin{eqnarray}
\Xi ^{L}(k_{y},s)\approx -\ddot{X}^{L}(k_{y},s).
\end{eqnarray}
On the other hand, when the correlation function $\dot{X}^{L}(k_{y},s)$ is
very small, the memory function (\ref{eqn:shearmemory}) is then expanded as
follows, 
\begin{eqnarray}
\Xi ^{L}(k_{y},s)=-\ddot{X}^{L}(k_{y},s)+\ddot{X}^{L}(k_{y},s)\dot{X}%
^{L}(k_{y},s)-\cdots .
\end{eqnarray}
That is, the correlation function $\dot{X}^{L}(k_{y},s)$ represents the
correction to the Q approximation.

As a matter of fact, in the Q approximation, we can derive the relativistic
Navier-Stokes equation and then our formula reproduces the GKN formula of
the shear viscosity coefficient. First of all, the correlation function $%
\ddot{X}(k_{y},t)$ is rewritten as 
\begin{eqnarray}
\ddot{X}^{L}(\mathbf{k},s) &=&\int_{0}^{\infty }dte^{-st}\theta (t)\frac{1}{%
10}\int d^{3}\mathbf{x}d^{3}\mathbf{x}_{1}e^{-i\mathbf{kx}}\mathbf{k}%
^{2}(\pi ^{\alpha \beta }(\mathbf{x},t),\pi _{\alpha \beta }(\mathbf{x}%
_{1},0))(T^{0x}(\mathbf{x}_{1},0),T^{0x}(\mathbf{0},0))^{-1}  \nonumber \\
&=&-\frac{1}{10\beta }\int_{0}^{\infty }dt\int d^{3}\mathbf{x}d^{3}\mathbf{x}%
_{1}e^{-st-i\mathbf{kx}}\mathbf{k}^{2}\int_{0}^{\infty }d\tau \langle \pi
^{\alpha \beta }(\mathbf{x},\tau )\pi _{\alpha \beta }(\mathbf{x}%
_{1},0)\rangle _{ret}(T^{0x}(\mathbf{x}_{1},0),T^{0x}(\mathbf{0},0))^{-1},
\end{eqnarray}
where 
\begin{eqnarray}
\langle \pi ^{\alpha \beta }(\mathbf{x},t)\pi _{\alpha \beta }(\mathbf{x}%
_{1},s)\rangle _{ret}=-i\theta (t-s)\langle \lbrack \pi ^{\alpha \beta }(%
\mathbf{x},t),\pi _{\alpha \beta }(\mathbf{x}_{1},s)]\rangle _{eq}.
\end{eqnarray}
In this derivation, we used the relation \cite{hosoya} 
\begin{eqnarray}
(\pi _{\mu \nu },\pi _{\rho \sigma })=\frac{L_{\pi }}{2}\left( \Delta _{\mu
\rho }\Delta _{\nu \sigma }-\Delta _{\mu \sigma }\Delta _{\nu \rho }-\frac{2%
}{3}\Delta _{\mu \nu }\Delta _{\rho \sigma }\right) ,
\end{eqnarray}
where $L_{\pi }$ is a scalar function and 
\begin{eqnarray}
\Delta _{\mu \nu }=g_{\mu \nu }-u_{\mu }u_{\nu },
\end{eqnarray}
with $u_{\mu }$ the four-velocity of the fluid in the Landau frame. The
correlation function $\ddot{X}^{L}(\mathbf{k},-i\omega +\epsilon )$ is real
in the low momentum limit. Then, the frequency function vanishes and the
equation of the energy-momentum tensor (\ref{T0x(yk)}) is given by the
linearized relativistic Navier-Stokes equation, 
\begin{eqnarray}
\frac{\partial }{\partial t}T^{0x}-\eta ^{NS}k_{y}^{2}T^{0x}=0,
\end{eqnarray}
where the Navier-Stokes shear viscosity coefficient is 
\begin{eqnarray}
\lim_{\mathbf{k}\rightarrow 0}\frac{1}{\tau _{\mathbf{k}}}=-\eta ^{NS} 
\mathbf{k}^{2}.  \label{eqn:NSlimit}
\end{eqnarray}
By using the expression of $\tau _{\mathbf{k}}$, the Navier-Stokes shear
viscosity coefficient is expressed by using the time correlation function as
follows, 
\begin{eqnarray}
\eta ^{NS}=\int_{-\infty }^{\infty }dt\theta (t)\frac{1}{10}\int d^{3}%
\mathbf{x}_{1}(\pi ^{\alpha \beta }(\mathbf{x},t),\pi _{\alpha \beta }(%
\mathbf{x}_{1},0))(T^{0x}(\mathbf{x}_{1},0),T^{0x}(\mathbf{0},0))^{-1}.
\end{eqnarray}
Except for the normalization factor $(T^{0x}(\mathbf{x}_{1},0),T^{0x}(%
\mathbf{0},0))^{-1}$, this expression is nothing but the GKN formula of the
shear viscosity coefficient \cite{hosoya,footnote1}. That is, our new
formula can reproduce the result of the GKN formula when the correlation
function $\dot{X}^{L}(k_{y},s)$ disappears in the low momentum limit. In
this sense, our formula is the generalization of the GKN formula.

As was pointed out, the vanishing $\dot{X}^L (k_y,s)$ corresponds to the Q
approximation. So far, because the exact expression of the memory function (%
\ref{eqn:mf}) was not known, we could not discuss whether the Q
approximation is applicable in the low momentum limit or not. Now the
validity of the Q approximation can be quantitatively estimated by
calculating the correlation function $\dot{X}^L (k_y,s)$. In fact, it is
already known that there are examples where the Q approximation cannot be
applicable \cite{koide6,km}

\section{Mode-Coupling Theory of Density Fluctuations}

\label{chap:n2mct}

So far, we discussed the simplest case where the system has only one gross
variable. In this section, we will consider a more complex case where we
need two gross variables to define the Mori projection operator. Such a
situation will occur, for example, in a glass dynamics \cite{mct}. Glass is
a high density system and a particle is thickly surrounded by other
particles. The energy and momentum of the particles are continuously
exchanged by collisions. However, it is difficult for particles to move away
from the initial position because the space around has already occupied by
others. This is called jamming. In a glass dynamics, we usually choose the
gross variables as the fluctuations of density of particles and the
corresponding current.

We consider a classical $N$-particle system, where the Hamiltonian is given
by \cite{mct} 
\begin{eqnarray}
H=\sum_{i}\frac{\mathbf{p}_{i}^{2}}{2m}+\frac{1}{2}\sum_{i\neq j}\phi (%
\mathbf{r}_{ij}).
\end{eqnarray}
Then, the corresponding Liouville operator is given by 
\begin{eqnarray}
iL=\frac{1}{m}\sum_{i}\left( \mathbf{p}_{i}\cdot \frac{\partial }{\partial 
\mathbf{r}_{i}}\right) -\sum_{i\neq j}\left( \frac{\partial \phi (\mathbf{r}%
_{ij})}{\partial \mathbf{r}_{ij}}\cdot \frac{\partial }{\partial \mathbf{p}%
_{i}}\right) .
\end{eqnarray}
In this case, the Fourier transform of the fluctuations of the particle
number density $\rho _{k}\left( t\right) $ is given by 
\begin{eqnarray}
\delta \rho _{\mathbf{k}}(t)=\sum_{i}e^{i\mathbf{k\cdot r}_{i}(t)}-(2\pi
)^{3}\rho \delta (\mathbf{k}),
\end{eqnarray}
where $\rho =N/V$ and $V$ is the volume of our system. In this system, the
number of particle is a conserved quantity and the fluctuations of the
density should satisfy the equation of continuity, 
\begin{eqnarray}
\delta \dot{\rho}_{\mathbf{k}}(t)=i|\mathbf{k}|j_{\mathbf{k}}(t).
\label{eqn:mcteoc}
\end{eqnarray}
Here, we define the current, 
\begin{eqnarray}
j_{\mathbf{k}}=\frac{1}{m}\sum_{i}(\hat{\mathbf{k}}\cdot \mathbf{p}_{i})e^{i%
\mathbf{k\cdot r}_{i}},
\end{eqnarray}
where $\hat{k}=\mathbf{k}/|\mathbf{k}|$. Then, we choose the set of the
gross variables as follows, 
\begin{eqnarray}
\mathbf{A}=\left( 
\begin{array}{c}
\delta \rho _{\mathbf{k}} \\ 
j_{\mathbf{k}}%
\end{array}%
\right) .
\end{eqnarray}
By substituting into Eq. (\ref{eqn:TC-1}), we obtain the evolution equation
of $\mathbf{A}$. We further multiply $\mathbf{A}^{\dagger }$ from the right
and take the thermal expectation value. Then the evolution equation of the
correlation function is given by, 
\begin{eqnarray}
\frac{\partial }{\partial t}\mathbf{C}(t)=i\Delta \mathbf{C}%
(t)-\int_{0}^{t}d\tau \mathbf{\Xi }(\tau )\mathbf{C}(t-\tau ),
\end{eqnarray}
where 
\begin{eqnarray}
\mathbf{C}(t) &=&\langle \mathbf{A}(t)\mathbf{A}^{\dagger }\rangle _{eq} 
\nonumber \\
&=&\left( 
\begin{array}{cc}
\langle \delta \rho _{\mathbf{k}}(t)\delta \rho _{\mathbf{-k}}\rangle _{eq}
& \langle \delta \rho _{\mathbf{k}}(t)j_{\mathbf{-k}}\rangle _{eq} \\ 
\langle j_{\mathbf{k}}(t)\delta \rho _{\mathbf{-k}}\rangle _{eq} & \langle
j_{\mathbf{k}}(t)j_{\mathbf{-k}}\rangle _{eq}%
\end{array}%
\right) .
\end{eqnarray}%
Because of the equation of continuity (\ref{eqn:mcteoc}), these four
correlation functions are not independent. It should be noted that, in the
usual discussion of the glass dynamics, we do not consider the thermal
equilibrium environment discussed here and the calculation of the memory
function is more involved.

The coefficient of the streaming term is given by 
\begin{eqnarray}
i\Delta = \left( 
\begin{array}{cc}
0 & ik \\ 
\frac{ik}{m\beta S(\mathbf{k})} & 0%
\end{array}
\right),
\end{eqnarray}
where the static structure factor is $S(\mathbf{k}) = \frac{1}{N}\langle
\rho_{\mathbf{-k}}(0) \rho_{\mathbf{k}}(0) \rangle$. By using Eq. (\ref%
{eqn:mf}), we found that the upper components of the matrix of the memory
function vanish, 
\begin{eqnarray}
\mathbf{\Xi}^L(s) = \left( 
\begin{array}{cc}
0 & 0 \\ 
\mathbf{\Xi}^L_{21}(s) & \mathbf{\Xi}^L_{22}(s)%
\end{array}
\right).
\end{eqnarray}
Because of $(\xi_i(t),A^{*}_j)=0$, the noise term disappears.

We will concentrate on the element in the lower column of the matrix. Then,
we can obtain the following two equations, 
\begin{eqnarray}
\frac{\partial^2}{\partial t^2}F(k,t) + \frac{k^2}{m\beta S(\mathbf{k})}
F(k,t) + \int^t_0 d\tau \left( \mathbf{\Xi}_{21}(\tau) (ik) F(k,t-\tau) + 
\mathbf{\Xi}_{22}(\tau) \frac{\partial}{\partial t} F(k,t-\tau) \right) &=&
0.  \label{eqn:F} \\
\frac{\partial^3}{\partial t^3}F(k,t) + \frac{k^2}{m\beta S(\mathbf{k})}%
\frac{\partial}{\partial t}F(k,t) + \int^t_0 d\tau \left( \mathbf{\Xi}%
_{21}(\tau) (ik) \frac{\partial}{\partial t} F(k,t-\tau) + \mathbf{\Xi}%
_{22}(\tau) \frac{\partial^2}{\partial t^2} F(k,t-\tau) \right) &=& 0.
\end{eqnarray}
Here, we introduce the following function, 
\begin{eqnarray}
F(k,t) = \frac{1}{N}\langle \delta \rho_{\mathbf{k}} (t) \delta \rho_{%
\mathbf{-k}} \rangle_{eq}.
\end{eqnarray}

From the consistency of the two equations, one can find that $\mathbf{\Xi}%
_{21}(\tau)$ should vanishes. As a matter of fact, this is shown by using
the following exact relation, 
\begin{eqnarray}
(\xi_i(t),\xi_j) = (\xi_j(-t),\xi_i)^* = [\mathbf{\Xi}(t)\cdot (\mathbf{A},%
\mathbf{A}^{\dagger})]_{ij}.
\end{eqnarray}
This relates the memory function and the noise term and is called the
fluctuation-dissipation theorem of second kind \cite{kubo}. From this
relation, one can show that $\mathbf{\Xi}_{21}(t)$ also should disappear
when $\mathbf{\Xi}_{12}(t)$ vanishes. The correlation functions, then,
should satisfy the following relation, 
\begin{eqnarray}
\mathbf{\Xi}_{12}^L (s) \propto ( \ddot{\mathbf{X}}^L_{21}(s) - [\dot{%
\mathbf{X}}(0)\dot{X}^L(s)]_{21} )(1+\dot{\mathbf{X}}^L_{22}(s)) - ( \ddot{%
\mathbf{X}}^L_{22}(s) - [\dot{\mathbf{X}}(0)\dot{X}^L(s)]_{22} ) \dot{%
\mathbf{X}}^L_{21}(s) = 0.
\end{eqnarray}
By using this relation, we can simplify the remaining memory term, 
\begin{eqnarray}
\mathbf{\Xi}_{22}^L (s) &=& -\frac{\ddot{\mathbf{X}}^L_{22}(s) - [\dot{%
\mathbf{X}}(0)\dot{\mathbf{X}}^L(s)]_{22}}{1+\dot{\mathbf{X}}^L_{22}(s)} 
\nonumber \\
&=& -\frac{s^2 F^L(k,s) - s S(\mathbf{k}) + k^2 F^L(k,s)/(m\beta S(\mathbf{k}%
) )} {s F^L(k,s) - S(\mathbf{k}) }.  \label{eqn:Xi22}
\end{eqnarray}
This is the result obtained by using our expression of the memory function (%
\ref{eqn:mf}).

In the case discussed here, however, all correlation functions are expressed
by the unique correlation function $F(k,t)$ and we do not need to use Eq. (%
\ref{eqn:mf}) to calculate the memory function. From Eq. (\ref{eqn:F}), the
Laplace transform of the equation is 
\begin{eqnarray}
s^2 F^L(k,s) - s F(k,0) - \dot{F}(k,0) = - \frac{k^2}{m \beta S(\mathbf{k})}%
F^L (k,s) - \mathbf{\Xi}_{22}^L (s) (s F^L(k,s) - F(k,0)).
\end{eqnarray}
The expression of the memory function that is obtained by solving the
equation above is same as Eq. (\ref{eqn:Xi22}). This means the consistency
of our formula.

\section{Shear Viscosity of Causal Dissipative Hydrodynamics in n=2 Form}

\label{chap:n2shear}

In the previous derivation of the causal shear viscosity, we assumed that
the macroscopic motion can be projected onto the space spanned by the unique
gross variable $T^{0x}(k_y)$. If this assumption is correct, the memory
function converges to a constant rapidly and we can define the causal shear
viscosity coefficient and the relaxation time as was done in section \ref%
{chap:n1shear}. The memory functions are calculated so far for several
examples and the behaviors are consistent with this assumption.

However, there is another suggestion for the definition of the projection
operator. In the wake of the discussion of the extended thermodynamics \cite%
{jou1}, Ichiyanagi proposed that the Mori projection operator should be
defined by using not only usual hydrodynamic variables but also the
corresponding currents \cite{ichi}, although any calculable formula was not
given. In this section, we rederive the formula for the causal shear
viscosity coefficient following his idea.

The set of the gross variables are given by 
\begin{eqnarray}
\mathbf{A} = \left( 
\begin{array}{c}
T^{0x}(k_y) \\ 
T^{yx}(k_y)%
\end{array}
\right).
\end{eqnarray}
By substituting into Eq. (\ref{eqn:TC-1}), we have 
\begin{eqnarray}
\frac{\partial}{\partial t}T^{0x}(k_y,t) &=& -ik_y T^{yx}(k_y,t),
\label{eqn:1} \\
\frac{\partial}{\partial t}T^{yx}(k_y,t) &=& -ik_y R_{k_y} T^{0x}(k_y,t) -
\int^t_0 d\tau \mathbf{\Xi}_{22}(\tau) T^{yx}(k_y,t-\tau),  \label{eqn:2}
\end{eqnarray}
where 
\begin{eqnarray}
R_{k_y} = (T^{yx}(k_y),T^{yx}(-k_y)) (T^{0x}(k_y),T^{0x}(-k_y))^{-1}.
\end{eqnarray}
The memory function is given by 
\begin{eqnarray}
\mathbf{\Xi}^L_{22}(s) &=& -\frac{\ddot{\mathbf{X}}^L_{22}(s) + ik_y R_{k_y} 
\dot{\mathbf{X}}^L_{12}(s)}{1+\dot{\mathbf{X}}^L_{22}(s)}  \nonumber \\
&=& -\frac{\ddot{\mathbf{X}}^L_{11}(s) + \ddot{\mathbf{X}}^L_{22}(s) }{1+%
\dot{\mathbf{X}}^L_{22}(s)}.
\end{eqnarray}
with the Laplace transforms of the following correlations, 
\begin{eqnarray}
\mathbf{X}(t) &=& \left( 
\begin{array}{cc}
( T^{0x}(k_y,t),T^{0x}(-k_y)) & ( T^{0x}(k_y,t),T^{yx}(-k_y)) \\ 
( T^{yx}(k_y,t),T^{0x}(-k_y)) & ( T^{yx}(k_y,t),T^{yx}(-k_y))%
\end{array}
\right) \left( 
\begin{array}{cc}
(T^{0x}(k_y),T^{0x}(-k_y))^{-1} & 0 \\ 
0 & (T^{yx}(k_y),T^{yx}(-k_y))^{-1}%
\end{array}
\right),  \nonumber \\
\dot{\mathbf{X}}(t) &=& \frac{\partial}{\partial t}\mathbf{X}(t), \\
\ddot{\mathbf{X}}(t) &=& \frac{\partial^2}{\partial t^2}\mathbf{X}(t).
\end{eqnarray}
Here, we omitted the noise term.

It should be noted that we still define Kubo's canonical correlation by Eq. (%
\ref{eqn:kubo}), where the expectation value is calculated by the usual
thermal equilibrium state. This is different from the original idea of
Ichiyanagi and the extended thermodynamics, where the concept of the
thermodynamic variables are extended and hence we have to use a
non-equilibrium state to calculate the expectation value. However, when we
restrict ourselves to the non equilibrium states whose deviation from
equilibrium is still small, then the expectation values can be evaluated at
the equilibrium, since the effect of the non-equilibrium expectation should
be higher order and we assume that it is negligible.

Equation (\ref{eqn:1}) is the equation of continuity. If we can derive the
causal dissipative hydrodynamics from the Heisenberg equation of motion, Eq.
(\ref{eqn:2}) should be reduced to the telegraph equation, 
\begin{eqnarray}
\frac{\partial}{\partial t}T^{yx}(k_y,t) &=& -\frac{\eta^{NN}}{2\tau} (ik_y)
u^x(k_y,t) - \frac{1}{\tau}T^{yx}(k_y,t).
\end{eqnarray}
It should be noted that when we combine this equation with the equation of
continuity, we can reproduce Eq. (\ref{eqn:c-hydro}). To obtain the
telegraph equation from Eq. (\ref{eqn:2}), we assume that the memory
function $\mathbf{\Xi}_{22}$ is Markovian; the memory function quickly
vanishes with time, 
\begin{eqnarray}
\mathbf{\Xi}_{22}(t) = \frac{2}{\tau_{k_y}}\delta(t),
\end{eqnarray}
where 
\begin{eqnarray}
\frac{1}{\tau_{k_y}} = \int^{\infty}_{0}d\tau \mathbf{\Xi}_{22}.
\end{eqnarray}
By substituting them into Eq. (\ref{eqn:2}), we obtain 
\begin{eqnarray}
\frac{\partial}{\partial t}T^{yx}(k_y,t) &=& -ik_y w R_{k_y} u^x(k_y,t) - 
\frac{1}{\tau_{k_y}} T^{yx}(k_y,t).
\end{eqnarray}

By comparison with the telegraph equation, we identify the causal shear
viscosity coefficient and the relaxation time as follows, 
\begin{eqnarray}
\tau &=&\lim_{k_{y}\rightarrow 0}\tau _{k_{y}},  \label{eqn:rela2} \\
\eta ^{NN} &=&\lim_{k_{y}\rightarrow 0}2R_{k_{y}}\tau .  \label{eqn:shear2}
\end{eqnarray}

Note that these formula look different form those in the previous section,
formula (\ref{eqn:rela1}) and (\ref{eqn:shear1}). We know that the two
approaches ($n=1$ and $n=2$) shown in this paper are completely equivalent
if no approximation is introduced, that is, the coupled equation (\ref{eqn:1}%
) and (\ref{eqn:2}) gives exactly same result as Eq. (\ref{eqn:pro-hydro}).
We even checked the consistency of the two approaches by solving the coupled
harmonic oscillator model which is exactly solvable \cite{km}.

To use the formula (\ref{eqn:rela2}) and (\ref{eqn:shear2}), we have to
calculate three correlation functions, while we need two correlation
functions in the formula (\ref{eqn:rela1}) and (\ref{eqn:shear1}). Thus we
should usually use the formula (\ref{eqn:rela1}) and (\ref{eqn:shear1}) to
estimate the causal shear viscosity coefficient in causal dissipative
hydrodynamics. However, when the calculated memory function does not satisfy
the condition (\ref{eqn:condition}), we have to use the formula (\ref%
{eqn:rela2}) and (\ref{eqn:shear2}).

\section{Concluding remarks}

\label{chap:dis}

In this paper, we derived the general expression of the memory function
extending the result of \cite{koide1}. By using the expression, we define
the shear viscosity coefficient and the corresponding relaxation time of the
causal dissipative hydrodynamics. Our formula is the generalization of the
GKN formula because, when the Q approximation is justified in the low
momentum limit, the GKN formula is reproduced.

Phenomenologically, the causal hydrodynamics is derived by introducing the
memory effect to the relation between irreversible currents and
thermodynamic forces. Thus the vanishing relaxation time limit $\tau
\rightarrow 0$ corresponds to the limit of the Newtonian fluid and hence the
causal dissipative hydrodynamics is reduced to the relativistic
Navier-Stokes equation (the Landau-Lifshitz theory). Thus it is sometimes
expected that the causal shear viscosity coefficient is still approximately
given by the calculation of the GKN formula, when the relaxation time is not
large. However, this expectation is not trivial. As was discussed in this
paper, the new formula reduced to the GKN formula in the Q approximation. In
this limit, as is shown in Eq. (\ref{eqn:NSlimit}), the causal shear
viscosity coefficient $\eta^{NN}$ vanishes and the expression of the
relaxation time $\tau$ is reduced to that of the shear viscosity coefficient
in the GKN formula. That is, what is approximately given by the GKN formula
is not the causal shear viscosity coefficient but the relaxation time.

By using the idea of the AdS/CFT (anti-de Sitter/conformal field theory)
correspondence in the string theory, we can calculate the correlation
function of the energy-momentum tensor in $\mathcal{N}=4$ supersymmetric
Yang-Mills theory \cite{LowViscosity}. From this result, we obtain $\eta/s =
1/(4\pi)$ with $s$ being the entropy density and many people expect that
this gives the minimum of the shear viscosity coefficient of the
relativistic fluid of quarks and gluons. It should be, however, noted that
the $\eta$ here is not $\eta^{NN}$ but $\eta^{NS}$, that is, this discussion
is true only for Newtonian fluids, because, to derive the result, the
expression of the GKN formula of the shear viscosity coefficient is used.
Thus, when we discuss the causal dissipative hydrodynamics, we cannot use
this value as the limit of the causal shear viscosity coefficient. The lower
bound of the shear viscosity coefficient may exist even for the causal
hydrodynamics. This will be predicted by using our new formula instead of
the GKN formula.

There are several approaches to derive the relativistic hydrodynamics
consistent with causality. However, as far as we know, the telegraph
equation plays an essential role to solve the problem of acausality in all
theories, and the difference of the theories comes from the non-linear
terms. Thus the formula discussed here is applicable even for other causal
dissipative hydrodynmaics, the Israel-Stewart theory \cite{ii}.
See \cite{dkkm4}, for more discussions about the relationship between different theories
The effect of non-linearity, in general, can change the coefficients of the linear
terms. To discuss the effect of non-linearity to the transport coefficients,
we have to consider the non-linear response \cite{kg}. In the projection
operator method, this is implemented by generalizing the projection operator
including non-linear terms. However, the quantitative effect has not been
known so far. 

On the other hand, the telegraph equation may be not unique solution of the
problem of acausality in hydrodynamcis. For example, there are different
approaches to solve this problem in diffusion processes 
\cite{kampen}. 
However, to our best knowledge, there is no formulation of causal
dissipative hydrodynamics in these alternative scenarios. It should also
worth mentioning that we have not so far encountered any problem in
implementing numerical simulations of the causal dissipative hydrodynamics 
\cite{dkkm2,dkkm4}.

It should be mentioned that 
the projection operator approach discussed here and the usual linear response theory 
do not have explicit Lorentz covariance, because we introduce a thermal equilibrium 
background.
That is, when the transport coefficients of relativistic fluids are calculated, 
we assume the existence of a local rest frame where the dynamics of macroscopic 
quantities are determined in a non-relativistic way, together with an appropriate boundary condition.

\vspace{1cm}

T. Koide acknowledges helpful discussions with D. Hou. This work is
supported by CNPq and FAPERJ.

\appendix

\section{The Green-Kubo-Nakano formula}

\label{app:gkn}

In the appendix, we gives the short review of the GKN formula. As for the
calculations of the GKN formula for relativistic fluid, see, for example, 
\cite{defu} and references therein.

We consider the system whose Hamiltonian is given by $H$. By applying an
external force, the total Hamiltonian is changed from $H$ to $H+H_{ex}(t)$,
with 
\begin{eqnarray}
H_{ex}(t) = - A F(t),
\end{eqnarray}
where $A$ is an operator and $F(t)$ is the c-number external force.

We consider the current $J$ induced by the external force. From the linear
response theory, we obtain 
\begin{eqnarray}
\langle J \rangle = \int^t_{-\infty}ds \Psi(t-s)F(s),  \label{eqn:GKN_ori}
\end{eqnarray}
where the response function is given by 
\begin{eqnarray}
\Psi(t) = \int^{\beta}_0 d\lambda \langle \dot{A}(-i\lambda)J(t)
\rangle_{eq}.
\end{eqnarray}
This is the exact result in the sense of the linear approximation. This
formula, also, is called the GKN formula. However, in particular, when we
define transport coefficients of hydrodynamics, we do not use this
expression.

In these cases, first, we assume the linear relation between currents and
the external force, $J(t) = D_{GKN} F(t)$ with the transport coefficient $%
D_{GKN}$. The formula to define the expression $D_{GKN}$ is the GKN formula
which is discussed in this paper. For this, we see that we should ignore the
memory effect (time-convolution integral) in Eq. (\ref{eqn:GKN_ori}), 
\begin{eqnarray}
\langle J \rangle \approx \int^{\infty}_{0}ds \Psi(s)F(t)
\end{eqnarray}
Then the GKN formula is 
\begin{eqnarray}
D_{GKN} = \int^{\infty}_{0}ds \Psi(s).
\end{eqnarray}

Thus this formula is applicable only when there is a proportional relation
between a current and a force, like Newtonian fluids. This is the reason why
we cannot use the GKN formula to calculate the transport coefficients of the
causal dissipative hydrodynamics.

In principle, it is possible to derive the transport coefficients of the
causal dissipative hydrodynamics from Eq. (\ref{eqn:GKN_ori}). Instead of $%
J(t) = D_{GKN} F(t)$, we assume the following telegraph equation, 
\begin{eqnarray}
\partial_t J(t) = -\frac{1}{\tau_R}J(t) + \frac{D}{\tau_R}F(t).
\label{eqn:teleg}
\end{eqnarray}
From Eq. (\ref{eqn:GKN_ori}), we can derive the following equation, 
\begin{eqnarray}
\partial_t J(t) = \Psi(0)F(t) + \int^{\infty}_0 ds \partial_s \Psi(s)F(t).
\end{eqnarray}
In the second term, we ignore the time-convolution integral. We further
assume the GKN formula to reexpress the first term. Then we finally obtain 
\begin{eqnarray}
\partial_t J(t) = \frac{\Psi(0)}{D_{GKN}}J(t)+ \int^{\infty}_0 ds \partial_s
\Psi(s)F(t).
\end{eqnarray}
By comparing this equation with Eq. (\ref{eqn:teleg}), we can derive the
expression of $D$ and $\tau_R$.

Exactly speaking, we considered here the current induced by the external
force. However, the shear viscosity is induced not by the external force but
by the difference of the boundary conditions. Thus the discussion is not
applicable to the problems discussed in this paper.

\section{The derivation of Eq. (22)}

\label{app:2}

In this appendix, we derive Eq. (\ref{matrix-xi}). By using Eqs. (\ref{xi})
and (\ref{B}), we obtain 
\begin{eqnarray}
\mathbf{\Xi}_{ij}(t) = -\theta(t) \sum_{k}(iLe^{iLt}\mathcal{B}%
(t)QiLA_i,A^{\dagger}_k)(A,A^{\dagger})^{-1}_{kj}.
\end{eqnarray}
The first three terms can be calculated as follows, 
\begin{eqnarray}
&& (iLe^{iLt}QiLA_i,A^{\dagger}_j)(A_j,A_j^{\dagger})^{-1} = \ddot{\mathbf{X}%
}_{ij}(t) - [\dot{\mathbf{X}}(0)\dot{\mathbf{X}}(t)]_{ij}, \\
&& (iLe^{iLt}(-i)\int^{t}_{0}ds
e^{-iLs}PLe^{iLs}QiLA_i,A_j^{\dagger})(A_j,A_j^{\dagger})^{-1} =
-\int^{t}_{0}ds [ [\ddot{\mathbf{X}}(s)\dot{\mathbf{X}} (t-s)]_{ij} 
\nonumber \\
&& - [\dot{\mathbf{X}}(0)\dot{\mathbf{X}}(s)\dot{\mathbf{X}}(t-s)]_{ij} ], \\
&& \lefteqn{(iLe^{iLt}\int^{t}_{0}ds_1 e^{-iLs_1}PiLe^{iLs_1}
\int^{s_1}_{0}ds_2 e^{-iLs_2}PiLe^{iLs_2} QiLA_i, A_j^{\dagger})
(A_j,A_j^{\dagger})^{-1} }  \nonumber \\
&&= \int^{t}_{0}ds_1\int^{s_1}_{0}ds_2 [ [\ddot{\mathbf{X}}(s_2)(\dot{%
\mathbf{X}}(s_1-s_2)\dot{\mathbf{X}}(t-s_1)]_{ij}  \nonumber \\
&& - [ \dot{\mathbf{X}} (0)\dot{\mathbf{X}} (s_2) \dot{\mathbf{X}} (s_1-s_2)%
\dot{\mathbf{X}} (t-s_1) ]_{ij} ].
\end{eqnarray}
In short, the n-th order term is given by 
\begin{eqnarray}
\lefteqn{(iLe^{iLt}(-i)^n \int^{t}_{0}ds_1 \cdots \int^{s_{n-1}}_{0}ds_n 
\breve{L}^P(s_1)\cdots \breve{L}^P(s_n)
QiLA_i,A_j^{\dagger})(A_j,A_j^{\dagger})^{-1} } &&  \nonumber \\
&=& (-1)^n \int^{t}_{0}ds_1 \cdots ds_n [ \ddot{\mathbf{X}}(s_n)\dot{\mathbf{%
X}}(s_{n-1}-s_n)\cdots \dot{\mathbf{X}}(s_1 - s_2)\dot{\mathbf{X}}(t-s_1)
]_{ij}  \nonumber \\
&& - (-1)^n \int^{t}_{0}ds_1 \cdots \int^{s_{n-1}}_{0} ds_n [ \dot{\mathbf{X}%
s}(0)\dot{\mathbf{X}}(s_{n})\dot{\mathbf{X}}(s_{n-1}-s_n)\cdots \dot{\mathbf{%
X}}(s_1-s_2)\dot{\mathbf{X}}(t-s_1) ]_{ij} .
\end{eqnarray}
By using this result, we can calculate 
\begin{eqnarray}
\sum_j (iLQe^{iLQs}iLA_1,A^{\dagger}_j)(A_j,A_1^{\dagger})^{-1} &=& (\ddot{%
\mathbf{X}}(t))_{11} - (\dot{\mathbf{X}}(0)\dot{\mathbf{X}}(t))_{11} 
\nonumber \\
&& - \int^{t}_{0}d [ (\ddot{\mathbf{X}}(s)\dot{\mathbf{X}}(t-s))_{11} - (%
\dot{\mathbf{X}}(0)\dot{\mathbf{X}}(s)\dot{\mathbf{X}}(t-s))_{11} ] 
\nonumber \\
&& - \int^{t}_{0}d_1 \int^{s_1}_{0}ds_2 [ (\ddot{\mathbf{X}}(s_2)\dot{%
\mathbf{X}}(s_1 - s_2)\dot{\mathbf{X}}(t-s_1))_{11} - (\dot{\mathbf{X}}(0)%
\dot{\mathbf{X}}(s_2)\dot{\mathbf{X}}(s_1 - s_2)\dot{\mathbf{X}}%
(t-s_1))_{11} ]  \nonumber \\
&& + \cdots.
\end{eqnarray}
In short, the Laplace transform of Eq. (\ref{matrix-xi}) is given by 
\begin{eqnarray}
(\mathbf{\Xi}^L(s))_{11} = -\left( \ddot{\mathbf{X}}^L(s)\frac{1}{1+\dot{%
\mathbf{X}}^L(s)} - \dot{\mathbf{X}}(0)\dot{\mathbf{X}}^L(s)\frac{1}{1+\dot{%
\mathbf{X}}^L(s)} \right)_{11}.
\end{eqnarray}
The other components of the matrix are calculated in the same way.


\begin{thebibliography}{99}
\bibitem{footnote2} This means that the memory function still has finite
correlation after long time. This is a kind of long-time tail of memory
function. This gives rise to a trouble when we interpret the equation as the
Langevin equation because it is known that the memory function is given by
the time correlation of the noise (The fluctuation-dissipation theorem of
second kind). In this calculation, however, the noise term is just the
component orthogonal to the gross variables and ignored. Thus, we will not
discuss this problem in this paper. The physical meaning of the long-time
tail may come from the incompleteness of the definition of the Mori
projection operator. See, also the discussion in \cite{km}.
\bibitem{footnote1} Exactly speaking, there can exist the difference of the
factor between the result of the POM and the GKN formula, because the linear
response theory sometimes has an ambiguity for the definition of the
transport coefficients. This is clear in the case of the definition of the
diffusion coefficient. It is known that the POM can predict the correct
expression, although the additional discussion is necessary to fix the
proportional factor in the linear response theory \cite{kubo}.
\end{thebibliography}
\end{document}